\documentclass[aps,pra,showpacs]{revtex4}
\usepackage{dcolumn}
\begin{document} 
[Phys. Rev. A {\bf 82}, 013637 (2010)]
\title{
Entangled superfluids: condensate dynamics of the entangled
Bose-Einstein condensation }
\author{Yu Shi}
\email{yushi@fudan.edu.cn}
\affiliation{Department of Physics, Fudan
University, Shanghai 200433, China}
\begin{abstract}
We study the condensate dynamics of the so-called entangled
Bose-Einstein condensation (EBEC), which is the ground state of a
mixture of two species of pseudospin-$\frac{1}{2}$ atoms with
interspecies spin-exchange scattering in certain parameter regimes. EBEC leads to four
inter-dependent superfluid components, each corresponding to the
orbital wave function associated with a spin component of a species.
The four superflows have various counter-relations, and altogether
lead to a conserved total supercurrent and a conserved total spin
supercurrent. In the homogenous case, we also obtain the elementary
excitations due to variations of the single-particle orbital wave
functions,  by exactly solving the generalized time-dependent
Bogoliubov equations. There are three gapless Bogoliubov modes and
one Klein-Gordon-like gapped mode.  The origin of these excitations
are also discussed from the perspective of spontaneous breaking of
the symmetries possessed by the system.
\end{abstract}


\maketitle

\section{Introduction}

Bose-Einstein condensation (BEC) amplifies quantum mechanical
behavior of individual particles into macroscopic quantum phases. In
the simplest case, BEC occurs in a single-particle spatial or
orbital  state. Quantum features are more pronounced when there are
additional degrees of freedom. For example, BEC occurring in a
superposed single particle state leads to Josephson
effect~\cite{leggett}, while BEC of atoms with spin degree of
freedom leads to a spinor condensate~\cite{spinor}. For spin $F=1$,
in a mean field state~\cite{mf,pethick}, which is exact for a gas with
ferromagnetic spin exchange, BEC occurs in a single-particle
superposition of the thee hyperfine states. On the other hand, in the exact ground
state for a spin-1 gas with antiferromagnetic spin
exchange~\cite{exact,pethick,ks,al}, BEC occurs in a superposition
of two-particle state. As a further development in this perspective,
the so-called entangled BEC (EBEC), i.e. BEC occurring in an
interspecies entangled two-particle state, amplifies entanglement of
individual distinguishable particles into a macroscopic phase of a
many-particle systems~\cite{shi0}. EBEC was found to be the ground
state of a mixture of two species of pseudospin-$\frac{1}{2}$ atoms
in a considerable parameter regime~\cite{shi1,shi2}. It is entirely
different from the two-component BEC, which occurs in a gas of two kinds of atoms
distinguished by only one degree of freedom~\cite{two,pethick},
whose ground state is simply a direct product of the states of the two
kinds of atoms, each separately undergoing BEC. Just like  the
simplest BEC may be a source of coherent atoms, EBEC could be a
source of entangled atom pairs.

EBEC, in the case where the total number of atoms of each species is equal to
$N$, refers to the many-body ground state
\begin{equation} |G_{0}\rangle = \frac{1}{\sqrt{N+1}N!}
(a_{\uparrow}^{\dagger}b_{\downarrow}^{\dagger}-
a_{\downarrow}^{\dagger}b_{\uparrow}^{\dagger})^{N}|0\rangle,
\label{gs1}
\end{equation}
where $a_{\sigma}$ and $b_{\sigma}$ are, respectively, the
annihilation operators of the two species $a$ and $b$ for pseudospin
$\sigma$ ($\sigma = \uparrow, \downarrow$). In $|G_0\rangle$, BEC
occurs  in a two-particle state of a maximally entangled
interspecies pair \begin{equation} \eta(\mathbf{r}_a,\mathbf{r}_b) =
\frac{1}{\sqrt{2}}
[\phi_{a\uparrow}(\mathbf{r}_a)|\uparrow\nobreak\rangle_a
\phi_{b\downarrow}(\mathbf{r}_b)|\downarrow\nobreak\rangle_b
-\phi_{a\downarrow}(\mathbf{r}_a)|\downarrow\nobreak\rangle_a
\phi_{b\uparrow}(\mathbf{r}_b)|\uparrow\rangle_b], \end{equation}
and is thus called EBEC or BEC with an entangled order parameter  $\eta(\mathbf{r}_a,\mathbf{r}_b)$.
Here $\phi_{\alpha\sigma}$ ($\alpha=a,b$, $\sigma=\uparrow,
\downarrow$) is the single-particle orbital wave function for each spin
state of an atom of each species. In the most general case,
spin dependence of the potential $U_{\alpha\sigma}$ and
the scattering lengths, which determine the orbital wave functions,
lead to a kind of spin-orbit coupling. Consequently,
$\phi_{\alpha\uparrow} \neq \phi_{\alpha\downarrow}$, hence spin and orbital parts in
$\eta(\mathbf{r}_a,\mathbf{r}_b)$ cannot be factorized as an orbital
part and a spin part, i.e.  there
exists spin-orbit ``entanglement''  in $\eta$. If, however,
$\phi_{\alpha\uparrow} = \phi_{\alpha\downarrow} = \phi_{\alpha}$, as in the case where they are dominantly determined by the spin-independent part of the Hamiltonian,  then
$\eta(\mathbf{r}_a,\mathbf{r}_b) =\phi_{a}(\mathbf{r}_a)
\phi_{b}(\mathbf{r}_b) \frac{1}{\sqrt{2}}
(|\uparrow\nobreak\rangle_a |\downarrow\nobreak\rangle_b
-|\downarrow\nobreak\rangle_a |\uparrow\rangle_b)$, i.e. spin and orbit become
disentangled, as in the usual consideration in SU(2) model spin-1 model, consequently, the entanglement between $a$-atom and
$b$-atom becomes entirely spin entanglement.

A state analogous to $|G_0\rangle$ appears in a SU(2) symmetric model
of a single species of pseudospin-$\frac{1}{2}$ atoms~\cite{ks,al},
with the role of the two different species played by the
single-particle orbital ground state and the first excited state of a single species of atoms.
But there are also differences: (i) In the SU(2) model, the number
of atoms in the two single-particle orbital states are not
conserved, but are fixed by a measurement or controlled by a
microcanonical distribution, while in EBEC the number of atoms in
the two species are strictly conserved. (ii) In the SU(2) model, in which there is only a single species, the
occupation of two orbital modes, rather than a single mode,  is
due to the constraint of spin conservation in the cooling process,
which could be compromised  as the collision rate during
the evaporative cooling might depend on hyperfine spins, and there could
be atom loss. In EBEC, in contrast, two distinguishable species can
have small total spin, as distinguishable atoms are not subject to the constraint of Bose symmetry, consequently all the atoms of each species can
occupy the lowest orbital modes.  (iii)  In the SU(2) model, the analog of $|G_0\rangle$ is not the true ground state of
the system. In EBEC,  $|G_0\rangle$ is the true
ground state. (iv)
In the SU(2) model, the orbital identity of atoms,  unlike different species,  and the
correlation between identical particles in the two orbital modes, unlike  the entanglement between atoms of different
species, are lost after the atoms are taken out of the trap.

An important open issue about EBEC is its condensate dynamics, i.e. superfluid behavior determined by the orbital wave
functions and elementary excitations due to the fluctuations of the orbital
wave functions.  In this article, under the presumption that EBEC exists,  we study the condensate  dynamics of EBEC, based on
a generalized version of the time-dependent Gross-Pitaevskii (GP)
equations.  EBEC leads to four inter-dependent components of the
superfluid with a few counter-relations between each other. We also
study the elementary excitations by exactly solving a set of
generalized version of the Bogoliubov equations, as well as from the
perspective of symmetry breaking.

\section{Hamiltonian and the generalized Gross-Pitaevskii equations}

Consider a dilute gas of two species of atoms in a trap. Each atom
has an internal degree of freedom  represented as a
pseudospin-$\frac{1}{2}$.
The field theoretic
Hamiltonian is
\begin{equation}
{\cal H} = \sum_{\alpha\sigma}\int d^3 r
\psi_{\alpha\sigma}^{\dagger} h_{\alpha\sigma} \psi_{\alpha\sigma} +
\frac{1}{2}\sum_{\alpha\sigma\sigma'}g^{(\alpha\alpha)}_{\sigma\sigma'}
\int d^3r
\psi^{\dagger}_{\alpha\sigma}\psi^{\dagger}_{\alpha\sigma'}\psi_{\alpha\sigma'}\psi_{\alpha\sigma}
+ {\cal H}_{ab},\end{equation} with
\begin{equation}
{\cal H}_{ab} = \sum_{\sigma\sigma'}g^{(ab)}_{\sigma\sigma'} \int
d^3r
\psi^{\dagger}_{a\sigma}\psi^{\dagger}_{b\sigma'}\psi_{b\sigma'}
\psi_{a\sigma} + g_e\int d^3 r
(\psi^{\dagger}_{a\uparrow}\psi^{\dagger}_{b\downarrow}
\psi_{b\uparrow}\psi_{a\downarrow}
+\psi^{\dagger}_{a\downarrow}\psi^{\dagger}_{b\uparrow}
\psi_{b\downarrow}\psi_{a\uparrow}),
\end{equation}
where $\alpha = a, b$ represents the two species, $\sigma = \uparrow, \downarrow$ represents the two basis states of the pseudospin-$\frac{1}{2}$, $h_{\alpha\sigma} =
-\hbar^2\nabla_\alpha^2/2m_\alpha+U_{\alpha\sigma}$ is the single
particle Hamiltonian, $\psi_{\alpha\sigma}$ is the field operator of
species $\alpha$. The coefficients $g$'s are shorthands for
$g^{(\alpha\beta)}_{\sigma_1\sigma_2\sigma_3\sigma_4}\equiv
(2\pi\hbar^2\xi^{(\alpha\beta)}_{\sigma_1\sigma_2\sigma_3\sigma_4}/\mu_{\alpha\beta})$,
where $\xi^{(\alpha\beta)}_{\sigma_1\sigma_2\sigma_3\sigma_4}$ is
the scattering length for the scattering in which an $\alpha$-atom
flips from $\sigma_4$ to $\sigma_1$ while an $\beta$-atom flips from
$\sigma_3$ to $\sigma_2$,
$\mu_{\alpha\beta}=m_{\alpha}m_{\beta}/(m_{\alpha}+m_{\beta})$ is
the effective mass. For scattering lengths, we use the shorthands
$\xi^{(\alpha\alpha)}_{\sigma\sigma}\equiv
\xi^{(\alpha\alpha)}_{\sigma\sigma\sigma\sigma}$,
$\xi^{(\alpha\alpha)}_{\sigma\bar{\sigma}}\equiv
2\xi^{(\alpha\alpha)}_{\sigma\bar{\sigma}\bar{\sigma}\sigma} =
2\xi^{(\alpha\alpha)}_{\sigma\bar{\sigma}\sigma\bar{\sigma}}$ for
$\sigma\neq \bar{\sigma}$~\cite{leggett}, $\xi^{(ab)}_{\sigma\sigma'}\equiv
\xi^{(ab)}_{\sigma\sigma'\sigma'\sigma}$,
$\xi^{(ab)}_e=\xi^{(ab)}_{\uparrow\downarrow\uparrow\downarrow
}=\xi^{(ab)}_{\downarrow\uparrow\downarrow\uparrow }$~\cite{change}.
Correspondingly $g^{(\alpha\beta)}_{\sigma\sigma'} \equiv
2\pi\hbar^2\xi^{(\alpha\beta)}_{\sigma\sigma'}/\mu_{\alpha\beta}$,
that is, $g^{(\alpha\alpha)}_{\sigma\sigma}\equiv
g^{(\alpha\alpha)}_{\sigma\sigma\sigma\sigma}$,
$g^{(\alpha\alpha)}_{\sigma\bar{\sigma}}\equiv
2g^{(\alpha\alpha)}_{\sigma\bar{\sigma}\bar{\sigma}\sigma} =
2g^{(\alpha\alpha)}_{\bar{\sigma}\sigma\sigma\bar{\sigma}}$ for
$\sigma\neq \bar{\sigma}$, $g^{(ab)}_{\sigma\sigma'}\equiv
g^{(ab)}_{\sigma\sigma'\sigma'\sigma}$, $g_{e}\equiv
g^{(ab)}_{\uparrow\downarrow\uparrow\downarrow}
=g^{(ab)}_{\downarrow\uparrow\downarrow\uparrow}$~\cite{change}.

Under the single orbital mode approximation for each species,
for each atom of species $\alpha$ ($\alpha=a,b$)
and pseudospin $\sigma$ ($\sigma= \uparrow, \downarrow$), only the
single-particle spatial ground state
$\phi_{\alpha\sigma}(\mathbf{r})$ is occupied.
Therefore $\psi_{\alpha}=\alpha_{\sigma}\phi_{\alpha\sigma}$, where
$\alpha_{\sigma}$ is the annihilation operator corresponding  to the 
single-particle orbital wave function $\phi_{\alpha\sigma}$. Then
the many-body Hamiltonian can  be simplified as
\begin{widetext} \begin{equation} {\cal H} = \sum_{\alpha,\sigma}
f_{\alpha\sigma} N_{\alpha\sigma}+
\frac{1}{2}\sum_{\alpha,\sigma\sigma'}K^{(\alpha\alpha)}_{\sigma\sigma'}
N_{\alpha\sigma}N_{\alpha\sigma'}
+\sum_{\sigma\sigma'}K^{(ab)}_{\sigma\sigma'}N_{a\sigma}
N_{b\sigma'} + K_e (a^{\dagger}_{\uparrow}a_{\downarrow}
b^{\dagger}_{\downarrow}b_{\uparrow} +
a^{\dagger}_{\downarrow}a_{\uparrow}
b^{\dagger}_{\uparrow}b_{\downarrow}), \end{equation} \end{widetext}
where $N_{\alpha\sigma}=\alpha^{\dagger}_{\sigma}\alpha_{\sigma}$.
The total number of atoms of each species $N_{\alpha} =
N_{\alpha\uparrow} +N_{\alpha\downarrow}$ is conserved. The
coefficients $K$'s are shorthands for
\begin{equation} K^{(\alpha\beta)}_{\sigma_1\sigma_2\sigma_3\sigma_4}\equiv
g^{(\alpha\beta)}_{\sigma_1\sigma_2\sigma_3\sigma_4}\int
\phi_{\alpha\sigma_1}^*(\mathbf{r})\phi_{\beta\sigma_2}^*(\mathbf{r})
\phi_{\beta\sigma_3}(\mathbf{r})\phi_{\alpha\sigma_4}(\mathbf{r})
d^3r,  \label{k}
\end{equation}
that is,  $K^{(\alpha\alpha)}_{\sigma\sigma}\equiv
K^{(\alpha\alpha)}_{\sigma\sigma\sigma\sigma}$,
$K^{(\alpha\alpha)}_{\sigma\bar{\sigma}}\equiv
2K^{(\alpha\alpha)}_{\sigma\bar{\sigma}\bar{\sigma}\sigma} =
2K^{(\alpha\alpha)}_{\bar{\sigma}\sigma\sigma\bar{\sigma}}$ for
$\sigma\neq \bar{\sigma}$, $K^{(ab)}_{\sigma\sigma'}\equiv
K^{(ab)}_{\sigma\sigma'\sigma'\sigma}$, $K_{e}\equiv
K^{(ab)}_{\uparrow\downarrow\uparrow\downarrow}
=K^{(ab)}_{\downarrow\uparrow\downarrow\uparrow}$~\cite{change}.
$f_{\alpha\sigma} \equiv \epsilon_{\alpha\sigma}-
K^{(\alpha\alpha)}_{\sigma\sigma}/2,$ where $\epsilon_{\alpha\sigma}
= \int \phi_{\alpha\sigma}^*h_{\alpha\sigma} \phi_{\alpha\sigma}
d^3r $ is the single-particle energy.

When $N_a=N_b$, the ground state is exactly $|G_0\rangle$,  under
the following conditions, which ensure consistency of
simplifying the Hamiltonian to that of isotropic Heisenberg coupling
between two giant spins representing the two species: (i)
$U_{\alpha\uparrow}(\mathbf{r})=U_{\alpha\downarrow}(\mathbf{r})=U_{\alpha}(\mathbf{r})$.
(ii) The intraspecies scattering lengths satisfy
$\xi^{(\alpha\alpha)}_{\sigma\bar{\sigma}}=\xi^{(\alpha\alpha)}_{\sigma\sigma}
=\xi_{\alpha}$~\cite{corr}. (iii) The interspecies scattering
lengths satisfy the relations
$\xi^{(ab)}_{\uparrow\uparrow}=\xi^{(ab)}_{\downarrow\downarrow},$
denoted as $\xi^{(ab)}_s$, and
$\xi^{(ab)}_{\uparrow\downarrow}=\xi^{(ab)}_{\downarrow\uparrow}$,
denoted as $\xi^{(ab)}_d$, where the subscripts ``s'' and ``d''
stand for  ``same'' and ``different'', respectively.  (iv)
$\xi^{(ab)}_e = \xi^{(ab)}_s-\xi^{(ab)}_d$. Under these conditions,
$\mu_{\alpha\uparrow}=\mu_{\alpha\downarrow}$,
$\phi_{\alpha\uparrow} = \phi_{\alpha\downarrow}$ in the many-body
ground state. The four wave functions $\phi_{\alpha\sigma}$'s, of which the condensate wave function $\eta$ is built on, satisfy four generalized GP equations. It has also
been shown that in a considerable parameter regime, the ground state
approaches $|G_0\rangle$. The conditions (i) and (ii) are
also satisfied in the SU(2) symmetric model of a single species of
pseudospin-$\frac{1}{2}$ atoms.

In the thermodynamic limit, the energetic advantage of EBEC is lost to the stability of simple BEC, like other fragmented BEC. Hence EBEC should be realized in a mesoscopic scale with a finite number of atoms and a finite volume. We can obtain upper bounds on the number of atoms as the following. According to previous discussions~\cite{shi2}, the effect of symmetry-breaking perturbation, which causes the ground state to deviate from EBEC, tends to diminish when the volume remains finite while $\Delta < 2K_e$~\cite{change}, where $\Delta$ is the energy gap of the perturbed Hamiltonian, and is given by $\Delta = \sqrt{4K_e d}$, where $d \equiv  |J_z-2K_e-C_a-C_b| N^2 +|B_b-B_a|N$, $N = (N_a+N_b)/2$, where $J_z=K_{\uparrow\uparrow}^{(ab)}+K_{\downarrow\downarrow}^{(ab)}
-K_{\uparrow\downarrow}^{(ab)}-K_{\downarrow\uparrow}^{(ab)},$
$B_a=f_{a\uparrow}-f_{a\downarrow}+\frac{N_a}{2}(K_{\uparrow\uparrow}^{(aa)}
-K_{\downarrow\downarrow}^{(aa)})
+\frac{N_b}{2}(K_{\uparrow\uparrow}^{(ab)}
+K_{\uparrow\downarrow}^{(ab)} -K_{\downarrow\uparrow}^{(ab)}
-K_{\downarrow\downarrow}^{(ab)}),$
$B_b=f_{b\uparrow}-f_{b\downarrow}+\frac{N_b}{2}(K_{\uparrow\uparrow}^{(bb)}
-K_{\downarrow\downarrow}^{(bb)})+ \frac{N_a}{2}(K_{\uparrow\uparrow}^{(ab)}
+K_{\downarrow\uparrow}^{(ab)}
-K_{\uparrow\downarrow}^{(ab)}-K_{\downarrow\downarrow}^{(ab)})$, with  $f_{\alpha\sigma} = \epsilon_{\alpha\sigma} - K_{\sigma\sigma}^{\alpha\alpha}$,
$C_{\alpha} = \frac{1}{2}(K_{\uparrow\uparrow}^{(\alpha\alpha)}
+K_{\downarrow\downarrow}^{(\alpha\alpha)}
-K_{\uparrow\downarrow}^{(\alpha\alpha)}
-K_{\downarrow\uparrow}^{(\alpha\alpha)})$, ($\alpha = a, b$). Under the conditions given in the above paragraph, $\Delta \rightarrow 0$, hence EBEC is indeed the ground state.  With deviation from these conditions, we may use the requirement $\Delta < 2K_e$, i.e. $d < K_e$,  to derive a constraint on $N$ for the occurrence of EBEC, which turns out to be $N < (\sqrt{|B_b-B_a|^2+K_e |J_z-2K_e-C_a-C_b|}-|B_b-B_a|)/(2|J_z-2K_e-C_a-C_b|)$. Furthermore, a necessary condition for EBEC to occur is that $k_B T < E_1-E_0 $, where $E_1$ and $E_0$ are the energy of the first excited and the EBEC ground states. $E_1-E_0 \approx 2K_e$. According to (\ref{k}), $K_e =
g_e \int
\phi_{a\uparrow}^*(\mathbf{r})\phi_{b\downarrow}^*(\mathbf{r})
\phi_{b\uparrow}(\mathbf{r})\phi_{a\downarrow}(\mathbf{r})
d^3r$. To make a rough estimation using uniform wave functions, we have $K_e = g_e/\Omega$, where $\Omega$ is the volume of the gas. Hence we should have  $T <2 g_e/k_B\Omega$. Alternatively, for  atoms in a trap, we may roughly assume $\phi_{\alpha\sigma}  \approx (\frac{m\omega}{\pi\hbar})^{3/4}\exp (-\frac{m\omega r^2}{2\hbar})$, where it is assumed that the atoms of both species have equal mass, and that the trap is isotropic with frequency $\omega$.  Then $K_e \approx  g_e \sqrt{m\omega/(2\pi\hbar)}$. Hence we have $T < (g_e/k_B) \sqrt{2m\omega/(\pi\hbar)}$. BEC transition temperature can be roughly  estimated using one component of the gas with particle number $N/2$~\cite{shi2}. Hence $T_c \approx 3.31 \hbar^2 (N/\Omega)^{2/3}/m$ for a uniform gas or $T_c \approx 0.94 \hbar \omega N^{1/3}$ for a trapped gas. Moreover, combining the estimation of $T_c$ with the result derived from $k_B T < E_1-E_0$,   we obtain a further constraint on $N$, namely $N <(2g_e m)^{3/2}/ [(3.31 \hbar^2 k_B)^{3/2} \Omega^{1/2}]$ for a uniform gas, or $N < (2g_e)^3 m^{3/2}/[k_B^3 (2\pi \omega)^{3/2}\hbar ^{9/2}]$ for a trapped gas.

EBEC might be experimentally realizable by using two species of
spin-1 alkali atoms in an optical trap with hyperfine states
constrained in a two-dimensional subspace of $|\uparrow\rangle
\equiv |F=2,m_F=2\rangle$ and $|\downarrow\rangle \equiv
|F=1,m_F=1\rangle$~\cite{shi2}. In order that the spin-exchange
scattering is energetically guaranteed, we should have
$\epsilon_{a\uparrow}-\epsilon_{a\downarrow}
=\epsilon_{b\uparrow}-\epsilon_{b\downarrow}$~\cite{shi1}, where
$\epsilon_{\alpha\sigma}$ ($\alpha=a, b$) is the single particle
energy. However, for two different species of alkali atoms, the
hyperfine splitting is different. A method of overcoming this difficulty
is to apply a magnetic field such that
$\epsilon_{\alpha\uparrow}-\epsilon_{\alpha\downarrow}$, which  is
now the sum of the hyperfine splitting $A_{\alpha}$ plus the
difference in Zeeman shift of the two hyperfine states
$(2g_{\alpha,F=2}-g_{\alpha,F=1})\mu_B B$, is the same for the two
species. For an alkali atom, $g_{\alpha,F}=
[F(F+1)+J(J+1)-I_{\alpha}(I_{\alpha}+1)]/F(F+1)$, where $J=1/2$,
$I_{\alpha}$ is the nuclear spin of species $\alpha$. Consider
species $a$ to be $^{87}$Rb while species $b$ to be $^{85}$Rb. Then
$A_a=6835MHz$, $A_b=3036MHz$, $I_a = 3/2$, and $I_b=5/2$~\cite{pethick}.
It can be estimated that $B=0.325 T$.

If EBEC exists, then what about its physical properties? Here we focus on  its orbital dynamics, which is determined by the generalized time-dependent GP
equation
\begin{equation}\begin{array}{c}
i\hbar
\displaystyle\frac{\partial\phi_{\alpha\sigma}(\mathbf{r})}{\partial
t}=
\displaystyle\{-\frac{\hbar^2}{2m_{\alpha}}\nabla^2+U_{\alpha}(\mathbf{r})
+\frac{2(N-1)}{3}g^{(\alpha\alpha)}_{\sigma\sigma}
|\phi_{\alpha\sigma}(\mathbf{r})|^2+\frac{N-1}{3}g^{(\alpha\alpha)}_{\sigma
\bar{\sigma}} |\phi_{\alpha\bar{\sigma}}(\mathbf{r})|^2 \\
+\displaystyle\frac{N-1}{3} g^{(\alpha\bar{\alpha})}_{\sigma\sigma}
|\phi_{\bar{\alpha}\sigma}(\mathbf{r})|^2 +
\frac{2N+1}{3}g^{(\alpha\bar{\alpha})}_{\sigma\bar{\sigma}}
|\phi_{\bar{\alpha}\bar{\sigma}}(\mathbf{r})|^2 \}
\phi_{\alpha\sigma}(\mathbf{r})\\
-\displaystyle\frac{N+2}{3}g_e
\phi^*_{\bar{\alpha}\bar{\sigma}}(\mathbf{r})
\phi_{\bar{\alpha}\sigma}(\mathbf{r})\phi_{\alpha\bar{\sigma}}(\mathbf{r}),
\end{array} \label{gp}
\end{equation}
where $\bar{\alpha}\neq \alpha$ represents the species other than
species $\alpha$, and $\bar{\sigma}\neq \sigma$ represents the
pseudospin opposite to $\sigma$.

These time-dependent GP equations can
be obtained from the static GP equations~\cite{shi1}, by replacing
$\mu_{\alpha\sigma}$ as $i\hbar\partial/\partial t$, and can also be
justified by using  the action principle
\begin{equation}
\delta \int_{t_1}^{t_2} L
dt =0, \end{equation}
where
\begin{equation}
L = \int d\mathbf{r} \{ (i\hbar/2) \sum_{\alpha
\sigma} (\phi_{\alpha\sigma}^*\partial\phi_{\alpha\sigma}/\partial t
- \phi_{\alpha\sigma}\partial\phi_{\alpha\sigma}^*/\partial t) -
\langle G_0 |{\cal H}|G_0\rangle\}
\end{equation}
is the Lagrangian functional for
$|G_0\rangle$. The static GP-like equations have been obtained
minimization of the energy functional under $|G_0\rangle$. This
self-consistent determination of the equation of motion of the
orbital wave function actually underlies the derivation of the
simplest GP equation, for which the many-body ground state is $(1/\sqrt{N!})(a^{\dagger})^N|0\rangle$. This methodology  has also been used, e.g., by Ashhab and Leggett in studying the SU(2) model~\cite{al}.

In studying the orbital dynamics of the simplest BEC, it is assumed that the system remains in BEC though the condensate wave function is time-dependent. Likewise, in the present case of EBEC, the generalized time-dependent GP  equations (\ref{gp}) presumes that the system remains in the many-body ground states in the form of $|G_0\rangle$, though the corresponding orbital wave functions $\phi_{\alpha \sigma}$ is time dependent. In other words, the system remains as BEC though the orbital wave functions are time dependent. It is this kind of dynamics that corresponds to superfluidity and is considered here.

\section{Supercurrents and spin supercurrents}

The number density of species $\alpha$ with pseudospin $\sigma$ is
\begin{equation}n_{\alpha\sigma}=(N/2)\phi_{\alpha\sigma}^*\phi_{\alpha\sigma},\end{equation}
while the supercurrent is
\begin{equation}
\mathbf{J}_{\alpha\sigma} = \frac{\hbar}{2mi}\frac{N}{2}
(\phi_{\alpha\sigma}^*\nabla\phi_{\alpha\sigma}-
\nabla\phi_{\alpha\sigma}^*\phi_{\alpha\sigma})=n_{\alpha\sigma}\mathbf{v}_{\alpha\sigma},
\label{velocity} \end{equation} where $\mathbf{v}_{\alpha\sigma}$ is
the superfluid velocity of species $\alpha$ of pseudospin $\sigma$.

From the generalized time-dependent GP equation (\ref{gp}), we
obtain
\begin{equation} \frac{\partial
n_{\alpha\sigma}(\mathbf{r},t)}{\partial t} +\nabla \cdot
\mathbf{J}_{\alpha\sigma} (\mathbf{r},t) = S_{\alpha\sigma},
\label{numbera}
\end{equation}
with
\begin{equation}
S_{a\sigma}=-S_{b\sigma}=-S_{a\bar{\sigma}}=S_{b\bar{\sigma}}=
-\frac{2(N+2)g_e}{3\hbar} Im (\phi_{b\bar{\sigma}}^* \phi_{b\sigma}
\phi_{a\bar{\sigma}} \phi_{a\sigma}^*),  \label{equal}
\end{equation}
which is due to interspecies spin-exchange and acts as a source.
Thus the supercurrent is not conserved individually in each
pseudospin component of each species. Equation~(\ref{equal}) indicates
the counter relations between the two pseudospin components of a
same species, as well as those between two components
with a same pseudospin and of two different species.

The total supercurrent of each species is conserved: 
\begin{equation}
\frac{\partial(n_{\alpha\uparrow}+n_{\alpha\downarrow})}{\partial t}
+\nabla \cdot
(\mathbf{J}_{\alpha\uparrow}+\mathbf{J}_{\alpha\downarrow})=0.\end{equation}
So is the total supercurrent for each pseudospin, \begin{equation}
\frac{\partial(n_{a\sigma}+n_{b\sigma})}{\partial t} +\nabla \cdot
(\mathbf{J}_{a\sigma}+\mathbf{J}_{b\sigma})=0.\end{equation} Of
course, the total supercurrent of the four components is conserved:
\begin{equation} \frac{\partial}{\partial t}\sum_{\alpha,\sigma}
n_{\alpha\sigma} +\nabla \cdot
\sum_{\alpha,\sigma}\mathbf{J}_{\alpha\sigma}=0.\end{equation}

Furthermore, we can also define spin density and spin supercurrent
for each species, \begin{equation}n^s_{\alpha}= m_F(\alpha,\uparrow)
n_{\alpha\uparrow}+ m_F(\alpha,\downarrow)
n_{\alpha\downarrow},\end{equation}
\begin{equation}\mathbf{J}^s_{\alpha}= m_F(\alpha,\uparrow)
\mathbf{J}_{\alpha\uparrow}+ m_F(\alpha,\downarrow)
\mathbf{J}_{\alpha\downarrow},\end{equation} where
$m_F(\alpha,\sigma)$ denotes the hyperfine $z$ component represented
by pseudospin $\sigma$ for species $\alpha$.

The spin supercurrent of each species is not conserved: 
\begin{equation}\begin{array}{c} \displaystyle\frac{\partial
n^s_{\alpha}}{\partial t} +\nabla \cdot J^s_{\alpha}
=[(m_F(\alpha,\uparrow)-m_F(\alpha,\downarrow)] S_{\alpha\uparrow} .
\end{array}\end{equation}

If  $m_F(a,\sigma) = m_F(b,\sigma)$, then the total spin
supercurrent of the two species is conserved, \begin{equation}
\frac{\partial(n_{a}^s+n_{b}^s)}{\partial t} +\nabla \cdot
(J_a^s+J_b^s)=0.\end{equation}

Thus we have a conserved total supercurrent as well as a conserved
total spin supercurrent.

\section{Hydrodynamics}

In parallel to some discussions in Ref.~\cite{pethick}, here we derive some hydrodynamic equations from the generalized GP equations. Our discussion is restricted to zero temperature, hence does not require thermodynamic limit and also applies to a finite system in a trap.

With $\phi_{\alpha\sigma}= f_{\alpha\sigma}
e^{i\Phi_{\alpha\sigma}}, $ where the phase $\Phi_{\alpha\sigma}$ is
not singular, Eq. (\ref{gp}) yields
\begin{equation}
\frac{\partial f^2_{\alpha\sigma}}{\partial t} =
-\frac{\hbar}{m_{\alpha}} \nabla \cdot (f^2_{\alpha\sigma}  \nabla
\Phi_{\alpha\sigma}) -\frac{2(N+2)g_e}{3\hbar}
f_{\alpha\sigma}f_{\alpha\bar{\sigma}}
f_{\bar{\alpha}\sigma}f_{\bar{\alpha}\bar{\sigma}} \sin
(\Phi_{\bar{\alpha}\bar{\sigma}}+\Phi_{\alpha\sigma}-\Phi_{\bar{\alpha}\sigma}-
\Phi_{\alpha\bar{\sigma}}),
\end{equation} which is just the continuity equation expressed in terms
of amplitudes and phases, and
\begin{equation}
\begin{array}{c} -\hbar \displaystyle\frac{\partial
\Phi_{\alpha\sigma}}{\partial t} = -\frac{\hbar^2}{2m_{\alpha}
f_{\alpha\sigma}} \nabla^2 f_{\alpha\sigma} +\frac{m_{\alpha}}{2}
v_{\alpha\sigma}^2 +U_{\alpha\sigma} + \frac{2(N-1)}{3}
g_{\sigma\sigma}^{(\alpha\alpha)} f_{\alpha\sigma}^2 +  \\
\displaystyle\frac{(N-1)}{3} g_{\sigma\bar{\sigma}}^{(\alpha\alpha)}
f_{\alpha\bar{\sigma}}^2 + \frac{N-1}{3}
g_{\sigma\sigma}^{(\alpha\bar{\alpha})} f_{\bar{\alpha}\sigma}^2 +
\frac{2N+1}{3}
g_{\sigma\bar{\sigma}}^{(\alpha\bar{\alpha})} f_{\bar{\alpha}\bar{\sigma}}^2 \\
-\displaystyle\frac{N+2}{3} g_e\frac{ f_{\bar{\alpha}\bar{\sigma}}
f_{\bar{\alpha}\sigma} f_{\alpha\bar{\sigma}} } {f_{\alpha\sigma}}
\cos(\Phi_{\bar{\alpha}\sigma}+\Phi_{\alpha\bar{\sigma}}-
\Phi_{\bar{\alpha}\bar{\sigma}}-\Phi_{\alpha\sigma}).\end{array}
\label{phi}
\end{equation}
According to (\ref{velocity}),
\begin{equation}\mathbf{v}_{\alpha\sigma} =
\frac{\hbar}{m}\nabla\Phi_{\alpha\sigma},\end{equation} hence the
gradient of (\ref{phi}) becomes
\begin{equation} m_{\alpha} \frac{\partial
\mathbf{v}_{\alpha\sigma}}{\partial t} = - \nabla
(\tilde{\mu}_{\alpha\sigma} +\frac{1}{2} m_{\alpha}
v^2_{\alpha\sigma}), \label{mdvdt} \end{equation} where
$\tilde{\mu}_{\alpha\sigma} =-\frac{\hbar^2}{2m_{\alpha} \sqrt{
n_{\alpha\sigma}} }\nabla^2 \sqrt{n_{\alpha\sigma}}
+U_{\alpha\sigma} + \frac{2(N-1)}{3}
g_{\sigma\sigma}^{(\alpha\alpha)} n_{\alpha\sigma} + \frac{(N-1)}{3}
g_{\sigma\bar{\sigma}}^{(\alpha\alpha)} n_{\alpha\bar{\sigma}} +
\frac{(N-1)}{3} g_{\sigma\sigma}^{(\alpha\bar{\alpha})}
n_{\bar{\alpha}\sigma} + \frac{(2N+1)}{3}
g_{\sigma\bar{\sigma}}^{(\alpha\bar{\alpha})}
n_{\bar{\alpha}\bar{\sigma}}+ V^{(e)}_{\alpha\sigma},$ where the
third to sixth terms on right-hand side are mean-field interaction energies due
to scattering without spin exchange, while
\begin{equation}V^{(e)}_{\alpha\sigma}=\frac{N+2}{3}
g_e\sqrt{\frac{n_{\bar{\alpha}\bar{\sigma}} n_{\bar{\alpha}\sigma}
n_{\alpha\bar{\sigma}} }{n_{\alpha\sigma}}}
\cos(\Phi_{\bar{\alpha}\sigma}+\Phi_{\alpha\bar{\sigma}}-
\Phi_{\bar{\alpha}\bar{\sigma}}-\Phi_{\alpha\sigma})\end{equation}
is due to spin exchange scattering. $\nabla V^e_{\alpha\sigma} =
[(N+2)/3]g_e(\nabla\sqrt{ n_{\bar{\alpha}\bar{\sigma}}
n_{\bar{\alpha}\sigma} n_{\alpha\bar{\sigma}}/ n_{\alpha\sigma}})
\cos(\Phi_{\bar{\alpha}\sigma}+\Phi_{\alpha\bar{\sigma}}-
\Phi_{\bar{\alpha}\bar{\sigma}}-\Phi_{\alpha\sigma})-[(N+2)/(3\hbar)]
g_e\sqrt{ n_{\bar{\alpha}\bar{\sigma}} n_{\bar{\alpha}\sigma}
n_{\alpha\bar{\sigma}} /n_{\alpha\sigma}}
\sin(\Phi_{\bar{\alpha}\sigma}+\Phi_{\alpha\bar{\sigma}}-
\Phi_{\bar{\alpha}\bar{\sigma}}-\Phi_{\alpha\sigma})
[m_{\bar{\alpha}}(\mathbf{v}_{\bar{\alpha}\sigma}-
\mathbf{v}_{\bar{\alpha}\bar{\sigma}})-
m_{\alpha}(\mathbf{v}_{\alpha\sigma}-\mathbf{v}_{\alpha\bar{\sigma}})],$
implying the inter-dependence between superfluid velocities of
different components.

It is interesting that with the spin-exchange term, Eq. (\ref{phi})
can still be written as a generalized Josephson relation
\begin{equation} \frac{\partial \Phi_{\alpha\sigma}}{\partial t}
=-\frac{1}{\hbar} \frac{\delta E}{\delta
n_{\alpha\sigma}},\end{equation} where \begin{equation} E=
E_a+E_b+E_{ab},\end{equation}  with \begin{equation}E_{\alpha} =
\frac{N}{2}\int d\mathbf{r} \{ \sum_{\sigma}
[\frac{\hbar^2}{2m_{\alpha}} |\nabla \phi_{\alpha \sigma}|^2 +
U_{\alpha\sigma} + \frac{N-1}{3} g_{\sigma\sigma}^{(\alpha\alpha)}
|\phi_{\alpha\sigma}|^4] + \frac{N-1}{3}
g_{\uparrow\downarrow}^{(\alpha\alpha)}
|\phi_{\alpha\uparrow}|^2|\phi_{\alpha\downarrow}|^2\},\end{equation}
\begin{equation}\begin{array}{rl} E_{ab}= & \displaystyle\frac{N}{2} \int
d\mathbf{r} \{\sum_{\sigma} [\frac{N-1}{3} g_{\sigma\sigma}^{(ab)}
|\phi_{a\sigma}|^2|\phi_{b\sigma}|^2 + \frac{2N+1}{3}
g_{\sigma\bar{\sigma}}^{(ab)}
|\phi_{a\sigma}|^2 |\phi_{b\bar{\sigma}}|^2]  \\
&  -  \displaystyle\frac{N+2}{3}g_e (\phi_{b\downarrow}^*
\phi_{b\uparrow}\phi_{a\uparrow}^*\phi_{a\downarrow}+
\phi_{a\downarrow}^*
\phi_{a\uparrow}\phi_{b\uparrow}^*\phi_{b\downarrow})\}, \end{array}
\end{equation}In case $\Phi_{\alpha\sigma}$ is independent of position
or time, this leads to time-independent GP equations.

Like the usual condensate, Eq. (\ref{mdvdt}) can be expressed in
terms of pressure,
\begin{equation}
\frac{\partial \mathbf{v}_{\alpha\sigma} }{\partial t} =
-\frac{1}{m_{\alpha} n_{\alpha\sigma}}\nabla P_{\alpha\sigma} -
\nabla (\frac{v_{\alpha\sigma}^2}{2}) + \frac{1}{m_{\alpha}} \nabla
(\frac{\hbar^2}{2m_{\alpha}\sqrt{n_{\alpha\sigma}}} \nabla^2
\sqrt{n_{\alpha\sigma}} ) -\frac{1}{m_{\alpha}} \nabla
U_{\alpha\sigma} , \label{mdvdt2} \end{equation} where pressure
$P_{\alpha\sigma}$  is related to chemical potential through
\begin{equation}dP_{\alpha\sigma} = n_{\alpha\sigma}d\mu'_{\alpha\sigma},
\end{equation} where
$\mu'_{\alpha\sigma}=\tilde{\mu}_{\alpha\sigma}-U_{\alpha\sigma}$.
$\mu'_{\alpha\sigma}$ and thus $P_{\alpha\sigma}$ are contributed by
interspecies spin-exchange scattering.

\section{Elementary excitations}

The variation $\delta \phi_{\alpha\sigma}$, away from
$\phi_{\alpha\sigma}$ in the ground state, satisfies the equation
\begin{equation}
\begin{array}{c} i\hbar
\displaystyle\frac{\partial}{\partial t}\delta\phi_{\alpha\sigma} =
(-\frac{\hbar^2}{2 m_{\alpha}} \nabla^2 + U_{\alpha\sigma})
\delta\phi_{\alpha\sigma}+\frac{4(N-1)}{3}g^{(\alpha\alpha)}_{\sigma\sigma}|\phi_{\alpha\sigma}|^2
\delta \phi_{\alpha\sigma}
+\frac{2(N-1)}{3}g^{(\alpha\alpha)}_{\sigma\sigma}
\phi_{\alpha\sigma}^2 \delta \phi_{\alpha\sigma}^*\\ \displaystyle
+\frac{(N-1)}{3}g^{(\alpha\alpha)}_{\sigma\bar{\sigma}}
|\phi_{\alpha\bar{\sigma}}|^2 \delta \phi_{\alpha\sigma} +
\frac{(N-1)}{3}g^{(\alpha\alpha)}_{\sigma\bar{\sigma}}
\phi_{\alpha\bar{\sigma}}^* \phi_{\alpha\sigma}
\delta\phi_{\alpha\bar{\sigma}} +
\frac{(N-1)}{3}g^{(\alpha\alpha)}_{\sigma\bar{\sigma}}
\phi_{\alpha\bar{\sigma}} \phi_{\alpha\sigma}
\delta\phi_{\alpha\bar{\sigma}}^* \\
\displaystyle + \frac{(N-1)}{3}
g^{\alpha\bar{\alpha}}_{\sigma\sigma} |\phi_{\bar{\alpha}\sigma}|^2
\delta \phi_{\alpha\sigma} + \frac{(N-1)}{3}
g^{\alpha\bar{\alpha}}_{\sigma\sigma} \phi_{\bar{\alpha}\sigma}^*
\phi_{\alpha\sigma}\delta \phi_{\bar{\alpha}\sigma}
 + \frac{(N-1)}{3}g^{\alpha\bar{\alpha}}_{\sigma\sigma} \phi_{\bar{\alpha}\sigma}
\phi_{\alpha\sigma} \delta \phi_{\bar{\alpha}\sigma}^* \\
\displaystyle + \frac{(2N+1)}{3}
g^{(\alpha\bar{\alpha})}_{\sigma\bar{\sigma}}
|\phi_{\bar{\alpha}\bar{\sigma}}|^2 \delta \phi_{\alpha\sigma} +
\frac{(2N+1)}{3}g^{(\alpha\bar{\alpha})}_{\sigma\bar{\sigma}}
\phi_{\bar{\alpha}\bar{\sigma}}^* \phi_{\alpha\sigma}\delta
\phi_{\bar{\alpha}\bar{\sigma}} + \frac{(2N+1)}{3}
g^{(\alpha\bar{\alpha})}_{\sigma\bar{\sigma}}
\phi_{\bar{\alpha}\bar{\sigma}}
\phi_{\alpha\sigma} \delta \phi_{\bar{\alpha}\bar{\sigma}}^*\\
\displaystyle - \frac{(N+2)}{3} g_e \phi_{\bar{\alpha}\sigma}
\phi_{\alpha\bar{\sigma}}\delta \phi_{\bar{\alpha}\bar{\sigma}}^*-
\frac{(N+2)}{3} g_e \phi_{\bar{\alpha}\bar{\sigma}}^*
\phi_{a\bar{\sigma}}  \delta\phi_{\bar{\alpha}\sigma}-
\frac{(N+2)}{3} g_e
\phi_{\bar{\alpha}\bar{\sigma}}^*\phi_{\bar{\alpha}\sigma} \delta
\phi_{\alpha\bar{\sigma}},
\end{array}
\label{variation}
\end{equation}
where $\phi_{\alpha\sigma}$ refers to orbital wave functions in the
ground state.

For simplicity, here we only consider a homogeneous system, i.e.
$U_{\alpha\sigma}=0$. It is reasonable to  suppose that a trapping potential $U_{\alpha\sigma}$ would not change the essential physics. Furthermore, we assume the above-mentioned conditions
$g^{(\alpha\alpha)}_{\sigma\sigma}=g^{(\alpha\alpha)}_{\sigma\bar{\sigma}}
=g_{\alpha}$, $g^{(ab)}_{\sigma\sigma}=g_{s}$, and 
$g^{(ab)}_{\sigma\bar{\sigma}}=g_{d}$. Then the chemical potentials
$\mu_{\alpha\uparrow}=\mu_{\alpha\downarrow}$, which is equal to
\begin{equation}
\mu_{\alpha}=\frac{N-1}{\Omega}g_{\alpha}+\frac{N-1}{3\Omega}g_s
+\frac{2N+1}{3\Omega}g_d -\frac{N+2}{3\Omega}g_e.
\label{mua}\end{equation} Thus single particle wave function in the
ground state is $\phi_{\alpha\sigma} = e^{-i\mu_{\alpha}t/\hbar}
/\sqrt{\Omega}$, which is then substituted to Eq.~(\ref{variation}).

Even though
$\phi_{\alpha\uparrow}=\phi_{\alpha\uparrow}$, their variations
should still be considered respectively.
Since we are considering a uniform system, we may set
\begin{equation}\delta\phi_{\alpha\sigma}
=\frac{e^{-i\mu_{\alpha\sigma}t/\hbar}}{\sqrt{\Omega}}
[u_{\alpha\sigma}(q) e^{i(\mathbf{q}\cdot\mathbf{r}-\omega t)}-
v_{\alpha\sigma}^*(q) e^{-i(\mathbf{q}\cdot\mathbf{r}-\omega
t)}],\end{equation} Then Eq.~(\ref{variation}) yields
\begin{equation}
\begin{array}{c} \displaystyle \omega u_{\alpha\sigma} =
W u_{\alpha\sigma}-\frac{2(N-1)}{3\Omega} g_{\alpha} v_{\alpha\sigma} \\
\displaystyle
 + [\frac{N-1}{3\Omega}g_{\alpha}
-\frac{N+2}{3\Omega} g_e ] u_{\alpha\bar{\sigma}}
-\frac{N-1}{3\Omega} g_{\alpha} v_{\alpha\bar{\sigma}} \\
\displaystyle + [\frac{N-1}{3\Omega} g_{s} - \frac{N+2}{3\Omega}g_e]
u_{\bar{\alpha}\sigma} -\frac{N-1}{3\Omega} g_{s}
v_{\bar{\alpha}\sigma}
\\ \displaystyle + \frac{2N+1}{3\Omega} g_{d}
u_{\bar{\alpha}\bar{\sigma}} + [-\frac{2N+1}{3\Omega} g_{d} +
\frac{N+2}{3\Omega}g_e ] v_{\bar{\alpha}\bar{\sigma}}
\end{array} \label{ua} \end{equation}
and
\begin{equation}
\begin{array}{c} \displaystyle \omega v_{\alpha\sigma}
=\frac{2(N-1)}{3\Omega} g_{\alpha} u_{\alpha\sigma}
 -W v_{\alpha\sigma} \\\displaystyle
+\frac{N-1}{3\Omega} g_{\alpha} u_{\alpha\bar{\sigma}} +
[-\frac{N-1}{3\Omega}g_{\alpha} +\frac{N+2}{3\Omega} g_e ]
v_{\alpha\bar{\sigma}}
 \\ \displaystyle +\frac{N-1}{3\Omega} g_s
u_{\bar{\alpha}\sigma} + [-\frac{N-1}{3\Omega} g_s +
\frac{N+2}{3\Omega}g_e] v_{\bar{\alpha}\sigma}
\\ \displaystyle + [\frac{2N+1}{3\Omega}
g_d - \frac{N+2}{3\Omega}g_e ]
u_{\bar{\alpha}\bar{\sigma}}-\frac{2N+1}{3\Omega} g_d
v_{\bar{\alpha}\bar{\sigma}}.
\end{array} \label{va} \end{equation}
where
\begin{eqnarray}
W & = & -\mu_{\alpha}
+\frac{\hbar^2q^2}{2m_{\alpha}} + \frac{5(N-1)}{3\Omega}g_{\alpha}
+\frac{N-1}{3\Omega} g_{s} +\frac{2N+1}{3\Omega} g_{d} \\
 & = & \frac{\hbar^2q^2}{2m_{\alpha}} + \frac{2(N-1)}{3\Omega}g_{\alpha}
+\frac{N+2}{3\Omega} g_{e},
\end{eqnarray}
where the second equality is a consequence of (\ref{mua}).

Therefore we have  eight coupled equations, which can be written as
a matrix \begin{equation} A\mathbf{U} = \omega
\mathbf{U},\end{equation} where \begin{equation}\mathbf{U}\equiv
(u_{a\uparrow}, v_{a\uparrow}, u_{a\downarrow},
v_{a\downarrow},u_{b\uparrow}, v_{b\uparrow}, u_{b\downarrow},
v_{b\downarrow})^T,\end{equation} the matrix elements of $A$ can be
read from Eqs.(\ref{ua}) and (\ref{va}).

We obtain four pairs of eigenvalues $\omega=\pm E_q^{(j)}/\hbar$ ($j=1,
2, 3, 4$),  $E_q^{(j)}$ being the energy of the elementary excitations.
As usual, a positive eigenvalue $\omega$ corresponds to addition of
a quasiparticle with momentum $\hbar \mathbf{q}$  and removal of a
quasiparticle with zero-momentum, while  a negative eigenvalue $\omega$
corresponds to removal of a quasiparticle with momentum
$-\hbar\mathbf{q}$ and addition of a quasiparticle with zero-momentum.

$E_q^{(j)}$ ($j=1,
2, 3, 4$) can be obtained exactly.  The first  and second excitations
are Bogoliubov-like gapless modes, and can be written together as
\begin{equation}
\displaystyle E^{(j)}_q = \{\frac{X_a+X_b}{2} \pm
[(\frac{X_a-X_b}{2})^2 + \frac{4 D}{9\Omega^2} \epsilon_q^a
\epsilon_q^b]^{\frac{1}{2}}\}^{\frac{1}{2}},
\end{equation}
where $j=1, 2$ correspond to $+$ and $-$ in $\pm$ respectively,
$\epsilon_q^{\alpha}= \hbar^2 q^2/2m_{\alpha}$, $X_{\alpha} \equiv
\epsilon_q^{\alpha} [\epsilon_q^{\alpha}+\frac{2(N-1)}{\Omega} g_{\alpha}], $ $D
\equiv [(N+2)g_e - (N-1)g_s - (2N+1) g_d]^2$. In terms of $q$,
\begin{equation}
 E^{(j)}_q =\hbar q \{ (\frac{1}{8m_a^2}+\frac{1}{8m_b^2})\hbar^2q^2 +
 \frac{N-1}{2\Omega}(\frac{g_a}{m_a}+\frac{g_b}{m_b}) \pm
 [\{ (\frac{1}{8m_a^2}-\frac{1}{8m_b^2})\hbar^2q^2 +
 \frac{N-1}{2\Omega}(\frac{g_a}{m_a}-\frac{g_b}{m_b})\}^2
 + \frac{D}{9\Omega^2 m_a m_b} ]^{\frac{1}{2}}\}^{\frac{1}{2}},
 \label{eq}
\end{equation}

If $m_a = m_b =m$, then these two energy spectra reduce to
\begin{equation}
 E^{(j)}_q = E_q^0 [E_q^0+\Gamma^{(j)}]\}^{\frac{1}{2}},
\end{equation}
where  $E_q^0 = \hbar^2q^2/2m$, $\Gamma^{(j)} \equiv
(N-1)(g_a+g_b)/\Omega \pm \{(N-1)^2(g_a-g_b)^2+4D/9\}^{1/2}/\Omega.$

From (\ref{eq}), it can be seen that in the long wavelength limit $q
\rightarrow 0$, the excitation $E^{(j)}_q$ ($j=1,2$) becomes linear
\begin{equation}E^{(j)}_q \rightarrow s^{(j)} \hbar q,\end{equation}
with  the sound velocity
\begin{equation} s^{(j)} = \{
 \frac{N-1}{2\Omega}(\frac{g_a}{m_a}+\frac{g_b}{m_b}) \pm
 [\{
 \frac{N-1}{2\Omega}(\frac{g_a}{m_a}-\frac{g_b}{m_b})\}^2
 + \frac{D}{9\Omega^2 m_a m_b} ]^{\frac{1}{2}}\}^{\frac{1}{2}},
\end{equation} The existence of
$E^{(2)}_q$ for $q \rightarrow 0$ is subject to the condition
$(N-1)^2 g_a g_b \geq D/9$, or  $ g_a g_b \geq (g_e - g_s -2g_d)^2$
for $N \gg 1$.

In the short wavelength limit  $q \rightarrow \infty$,
\begin{equation}E^{(1)}_q\rightarrow  \epsilon_q^a, \mbox{ } \mbox{ }
E^{(2)}_q\rightarrow  \epsilon_q^b, \end{equation}
that is, they reduce to single particle excitations.

The third and fourth excitations can be written as
\begin{equation}
\displaystyle E^{(j)}_q = [ Z_q  \mp (Z_q^2 -
Y_q)^{\frac{1}{2}}]^{\frac{1}{2}},
\end{equation}
where $j=3, 4$ correspond to $-$ and $+$ in $\pm$ respectively, $Z_q
\equiv \epsilon_q^a [\epsilon_q^a/2 + (N-1)g_a/(3\Omega) + 2 (N+2)g_e/(3\Omega)] +
\epsilon_q^b [\epsilon_q^b/2 + (N-1)g_b/(3\Omega) + 2 (N+2)g_e/(3\Omega)] + R$,
where $R \equiv 2 (N+2)g_e[(N-1)(g_a + g_b - 2 g_s) +
2(2N+1) g_d + 2(N+2)g_e]/(9\Omega^2),$ $Y_q \equiv \epsilon_q^a \epsilon_q^b (
\epsilon_q^a \epsilon_q^b + F  ) + 4
\epsilon_q^a (N+2)g_e \{[(N-1)g_b + (N+2)g_e] \epsilon_q^a/(9\Omega^2) +2C/3\Omega
\} + 4 \epsilon_q^b (N+2)g_e \{[(N-1)g_a + (N+2)g_e] \epsilon_q^b/(9\Omega^2)
+2C/3\Omega \}$, where $F \equiv 2[(N-1)g_b + 2(N+2)g_e]\epsilon_q^a/(3\Omega) + 2[(N-1)g_a +
2(N+2)g_e]\epsilon_q^b/(3\Omega) + 8 (N+2)^2 g_e^2/(9\Omega^2) + 4 C$, with  $C \equiv
\{[(N+2)g_e+(N-1)g_a][(N+2)g_e+(N-1)g_b] -
[(N-1)g_s-(2N+1)g_d]^2\}/(9\Omega^2)$.

It is then clear that  as $q
\rightarrow 0$, $Z_q \rightarrow R$, $Y_q \rightarrow 0$, hence  $E_q^{(3)}$ is gapless while  $E_q^{(4)}$ is gapped
with the gap $E_{q=0}^{(4)} = \sqrt{2R}$, under the condition $R
>0$, that is,  $(N-1)(g_a+g_b-2g_s)+2(2N+1)g_d+2(N+2)g_e >0$, or
$g_a+g_b-2g_s+4g_d+2g_e > 0$ when $N \gg 1$.

In the long wavelength limit $q \rightarrow 0$,
\begin{equation}E^{(3)}_q\rightarrow s^{(3)} \hbar q,\end{equation}
with  the sound velocity
\begin{equation} s^{(3)} =
 [\frac{2(N+2)g_e C}{3\Omega R} (\frac{1}{m_a}+\frac{1}{m_b})]^{\frac{1}{2}}.\end{equation}
 $E_q^{(3)}$ being gapless is subject to the condition $C/R \geq 0$.   When $N \gg 1$,
$s^{(3)} = [(g_e+g_a)(g_e+g_b)-(g_s+2g_d)^2](1/m_a+1/m_b)
/[3(g_a+g_b-2g_s+4g_d+2g_e)\Omega]$.

In the long wavelength limit $q \rightarrow 0$,
\begin{eqnarray}
E^{(4)}_q & \rightarrow & \sqrt{\Delta^2 + c^2 \hbar^2 q^2}
\\
& \approx & \Delta +  \frac{\hbar^2 q^2}{2m_{eff}},
\end{eqnarray}
where \begin{equation} \Delta = \sqrt{2R}
\end{equation}
is the energy gap at $q=0$, similar to the energy due to the rest
mass of a relativistic particle,
\begin{equation} c^2 = \frac{N-1}{3\Omega}(\frac{g_a}{m_a} +
\frac{g_b}{m_b}) + \frac{2(N+2)}{3\Omega}(\frac{g_e}{m_a} +
\frac{g_e}{m_b})(1-\frac{C}{R})
\end{equation}
is a constant similar to the square of the speed of light for a
relativistic particle,
\begin{equation} m_{eff}=\frac{\Delta}{c^2}
\end{equation}
is the effective rest mass of the particle-like excitation near
$q=0$. Therefore, in the long wavelength limit,  while the three
excitations $E_q^{(1)}$, $E_q^{(2)}$ and $E_q^{(3)}$ behave like
massless particles,  the fourth excitation $E_q^{(4)}$ behaves like
a massive Klein-Gordon particle, under the condition $R
> 0$ and $c^2 \geq 0$.

In the short wavelength limit  $q \rightarrow \infty$,
\begin{equation}E^{(3)}_q\rightarrow  \epsilon_q^a, \mbox{ } \mbox{ }
E^{(4)}_q\rightarrow  \epsilon_q^b \end{equation} if $m_a \geq m_b$.
$E^{(3)}_q\rightarrow  \epsilon_q^b, \mbox{ } \mbox{ } E^{(4)}_q\rightarrow
\epsilon_q^a$ if $m_b \geq m_a$.

It is always the case that when
$q \rightarrow \infty$, the excitations reduce to the free particle
spectra $\epsilon_q^a$ and $\epsilon_q^b$, each being double degenerate as there
are two pseudospin states for each species. The effect of interspecies spin exchange is manifested as $q \rightarrow 0$, in both a nonzero sound velocity of the third excitation and the nonzero gap of the fourth excitation.

\section{Nature of symmetry breaking}

From the point of view of symmetry breaking, Bogoliubov modes are
Goldstone modes associated with $U(1)$  gauge symmetry breaking. In
our model, there are three conserved particle numbers $N_a$, $N_b$
and
$N_{a\uparrow}+N_{b\uparrow}-N_{a\downarrow}-N_{b\downarrow}=2S_z$,
hence there are three $U(1)$ symmetries, the breaking of which gives
rise to the three Bogoliubov modes. U(1) group is isomorphic to
SO(2). Indeed, the U(1) symmetry generated by
$N_{a\uparrow}+N_{b\uparrow}-N_{a\downarrow}-N_{b\downarrow}$ is
just the SO(2) symmetry generated by $S_z$.

It is not difficult to identify $E_q^{(1)}$ and $E_q^{(2)}$ as the
Goldstone modes associated $N_a$ and $N_b$, respectively. In fact,
they reduce to the spectra of $a$-atoms and $b$-atoms if there is no
interspecies scattering. $E_q^{(3)}$ is associated with
$N_{a\uparrow}+N_{b\uparrow}-N_{a\downarrow}-N_{b\downarrow}$. It is
interesting that at the long-wavelength limit, $E_q^{(3)}$ depends
only on the spin-exchange scattering.

The gapped mode $E_q^{(4)}$ is due to the spin exchange between the two species.  In fact, the gap vanishes if $g_e=0$. At the
isotropic parameter point, the Hamiltonian can be rewritten as
${\cal H} = 2K_e \mathbf {S}_{a }\cdot\mathbf{S}_{b}$, where
$S_{\alpha}$ is the total spin of $\alpha$-species.  Hence
\begin{equation}
{\cal H}
= 2 K_e S_{a }S_{b}\cos\theta, \label{dt}
\end{equation}
where $S_a =N_a/2$, $S_b =N_b/2$,
$\theta$ is the angle between $S_{a}$ and $S_{b}$. For the ground
state, and $\theta$ is uniquely $\pi$.    Around  $\theta =\pi$, the
Hamiltonian  is
\begin{equation}{\cal H}_e = 2K_e S_{a}S_{b}
[-1+(\delta\theta)^2],\end{equation} where $\delta\theta \equiv
\theta-\pi$. Therefore $\delta\theta$ is massive. This leads to the
gapped mode.

One can also consider this issue in terms of the phases $\Phi_{\alpha\sigma}$ of the four
components. In the absence of the interspecies spin-exchange part, ${\cal H}_e$, of the
Hamiltonian,  $N_{\alpha\sigma}$ would be conserved. Then the spontaneous breaking of these four
$U(1)$ symmetries  would give four phases $\Phi_{\alpha\sigma}$'s,
as well as four Goldstone modes, each corresponding to a combination
of the four phases. With interspecies spin exchange, the Hamiltonian imposes an extra constraint on
these four phases,  as
\begin{equation}
{\cal H} = 2 K_e
(n_{a\uparrow}n_{a\downarrow}n_{b\uparrow} n_{b\downarrow})^{1/2}
\cos(\Phi_{a\uparrow}+\Phi_{b\downarrow}
-\Phi_{a\downarrow}-\Phi_{b\uparrow}), \label{dp}
\end{equation}
which fixes one combination
of the four phases.  By comparison of (\ref{dt}) and (\ref{dp}), it can be identified that
\begin{equation}
\theta =\Phi_{a\uparrow}+\Phi_{b\downarrow}
-\Phi_{a\downarrow}-\Phi_{b\uparrow}. \end{equation} Around the
minimum of ${\cal H}_e$,
\begin{equation}{\cal H}_e = 2 K_e
(n_{a\uparrow}n_{a\downarrow}n_{b\uparrow} n_{b\downarrow})^{1/2}
[-1+(\delta \Phi)^2],\end{equation} where $\delta\Phi \equiv
\Phi_{a\uparrow}+\Phi_{b\downarrow}-\Phi_{a\downarrow}
-\Phi_{b\uparrow}-\pi$. Hence the mode corresponding to $\delta\Phi$
becomes gapped. This gapped mode corresponds to
the source term $S_{\alpha\sigma}$ in
Eq.~(\ref{numbera}). Therefore there are only three Goldstone modes
remained. The gapped mode is due to the fluctuation of $\theta =
\Phi_{a\uparrow}+\Phi_{b\downarrow}
-\Phi_{a\downarrow}-\Phi_{b\uparrow}$.

In the number conserved ground state $|G_0\rangle$, $\langle
a_{\uparrow}^{\dagger} a_{\uparrow}\rangle = \langle
a_{\downarrow}^{\dagger}a_{\downarrow}\rangle = N/2$, implying
off-diagonal long-range order in each pseudospin component of each
species. Equivalently,  in the language of gauge symmetry
breaking, $U(1)$  gauge symmetry for each pseudospin component of
each species is broken in the symmetry breaking ground state, i.e. $\langle
a_{\alpha\sigma}\rangle = \sqrt{N/2} e^{i\Phi_{\alpha\sigma}} \neq
0$.

On the other hand, EBEC means $\langle a_{\uparrow}b_{\downarrow}-
a_{\downarrow}b_{\uparrow}\rangle \neq 0$, as can be justified as follows. In the symmetry breaking
ground state, $\langle a_{\uparrow}b_{\downarrow}-
a_{\downarrow}b_{\uparrow}\rangle = (N/2)\{ \exp
[i(\Phi_{a\uparrow}+\Phi_{b\downarrow})]-\exp
[i(\Phi_{a\downarrow}+\Phi_{b\uparrow})]\}.$ As
$\Phi_{a\uparrow}+\Phi_{b\downarrow}-\Phi_{a\downarrow}
-\Phi_{b\uparrow}=\pi$ in the ground state, we have $\exp
[i(\Phi_{a\uparrow}+\Phi_{b\downarrow})]= -\exp
[i(\Phi_{a\downarrow}+\Phi_{b\uparrow})]$. Therefore $\langle
a_{\uparrow}b_{\downarrow}- a_{\downarrow}b_{\uparrow}\rangle = N
\exp [i(\Phi_{a\uparrow}+\Phi_{b\downarrow})]$. Therefore, differing
from the case of molecular BEC, we have $\langle a_{\sigma}\rangle
\neq 0$, $\langle b_{\sigma}\rangle \neq 0$, as well as $\langle
a_{\uparrow}b_{\downarrow}- a_{\downarrow}b_{\uparrow}\rangle \neq
0$, i.e. both atoms and the nonlocal interspecies entangled pairs
are Bose-condensed. These field theoretical discussions are all
consistent with the exact results in the particle-number conserved
ground state above.

\section{summary}

To summarize, we have studied condensate dynamics of EBEC, which
exhibits peculiar superfluidity. EBEC leads to various counter
relations among the four superfluid components. The total
supercurrent and the total spin supercurrent are conserved.  For the
homogeneous case, we have also studied the elementary excitations due
to fluctuations of the four orbital wave functions around those in
the ground state of the system. There are four modes, three of which
are Bogoliubov-like while another is gapped. The three Bogoliubov
modes correspond to the spontaneous breaking of the $U(1)$
symmetries associated with three conserved particle numbers, while
the gapped mode is associated with the spin exchange. Alternatively,
the excitations can be understood as that the interspecies spin exchange gives
mass to a certain combination of the four Bogoliubov modes
corresponding to the four components of the system, hence there are
only three gapless modes remaining.

The emergence of massless elementary excitations of superfluids have been regarded as a paradigm showing how effective theory emerges from physics above the ``Planckian'' scale~\cite{volovik}. In this perspective, our model provides a way generating massive Klein-Gordon particles.

\bigskip

I thank Nigel Cooper, Anatoly Kuklov,Tony Leggett, Qian Niu, Jinlong Wang and Rukuan Wu for useful discussions.
This work is supported by National Science Foundation of China
(Grant No. 10674030), Shuguang Project (Grant No. 07S402) and
Ministry of Science and Technology of China (Grant No.
2009CB929204).


\begin{thebibliography}{99}
\bibitem{leggett} A. J. Leggett, Rev. Mod.
Phys. {\bf 73}, 307 (2001).
\bibitem{spinor} J. Stenger {\it et al.},
Nature {\bf 396}, 345 (1998); H.-J. Miesner {\it et al.}, Phy. Rev.
Lett. {\bf 82}, 2228 (1999); D. M. Stamper-Kurn {\it et al.}, Phy.
Rev. Lett. {\bf 83}, 661 (1999); H. Schmaljohann {\it et al.}, Phy.
Rev. Lett. {\bf 92}, 040402 (2004); M. S. Chang {\it et al.}, Phy.
Rev. Lett. {\bf 92}, 140403 (2004); T. Kuwamoto {\it et al.}, Phy.
Rev. A {\bf 69}, 063604 (2004); M. S. Chang  {\it et al.}, Nature
Phys. {\bf 1}, 111 (2005); L. E. Sadler {\it et al.}, Nature {\bf
443}, 312 (2006).
\bibitem{mf} T.-L. Ho, Phys. Rev. Lett.
81, 742 (1998); T. Ohmi and K. Machida, J. Phys. Soc. Jpn. 67, 1822
(1998).
\bibitem{pethick} C. J. Pethick and H. Smith,
{\em Bose-Einstein Condensation in Dilute Gases}, 2nd Edition
(Cambridge University Press, Cam bridge, 2008).
\bibitem{exact} C. K. Law, H. Pu, and N. P. Bigelow, Phys. Rev. Lett.
{\bf 81}, 5257 (1998); M. Koashi and M. Ueda, Phy. Rev. Lett. {\bf
84}, 1066 (2000); T. L. Ho and S. K. Yip, Phy. Rev. Lett. {\bf 84},
4031 (2000).
\bibitem{ks} A. B. Kuklov and B. V. Svistunov, Phy. Rev. Lett. {\bf 89}, 170403 (2002).
\bibitem{al} S. Ashhab and A. J. Leggett, Phys. Rev. A {\bf 68}, 063612
(2003).
\bibitem{shi0} Y. Shi,  Int. J. Mod. Phys. B {\bf 15}, 3007 (2001).
\bibitem{shi1} Y. Shi and Q. Niu, Phy. Rev. Lett. {\bf 96}, 140401
(2006).
\bibitem{shi2} Y. Shi, EPL {\bf 86}, 60008 (2009).
\bibitem{two} T. L. Ho and V. B. Shenoy, Phy. Rev. Lett. {\bf 77}, 3276
(1996);  C. J. Myatt, {\it et al.}, Phy. Rev. Lett. {\bf 78}, 586
(1997);  D. S. Hall {\it et al.}, Phy. Rev. Lett. {\bf 81}, 1539
(1998); D. S. Hall {\it et al.}, Phy. Rev. Lett. {\bf 81}, 1543
(1998); G. Modugno  {\it et al.}, Phy. Rev. Lett. {\bf 89}, 190404
(2002); G. Thalhammer {\it et al.}, Phy. Rev. Lett. {\bf 100},
210402 (2008); S. B. Papp, J. M. Pino and C. E. Wieman, Phy. Rev.
Lett. {\bf 101}, 040402 (2008).
\bibitem{change}  $\xi_e^{(ab)}$ here is the same as that
in \cite{shi2}, and was not defined in \cite{shi1}. $g_e$ and $K_e$
here correspond, respectively, to $g_e/2$ and $K_e/2$ in
\cite{shi1,shi2}.
\bibitem{corr} There is an error on this point in Ref.~\cite{shi2}.
\bibitem{volovik} G. E. Volovik, {\em The Universe in a Helium Droplet} (Cambridge University Press, Cambridge, 2003).
\end{thebibliography}
\end{document}